\documentclass[conference]{IEEEtran} \usepackage{fixltx2e} \usepackage{cite}
\usepackage[pdftex]{graphicx} \usepackage[cmex10]{amsmath}
\usepackage[caption=false,font=normalsize]{subfig} \usepackage{url}
\usepackage{multirow} \usepackage[utf8x]{inputenc} \usepackage[T1]{fontenc}
\usepackage{cleveref} \usepackage{booktabs}
\usepackage[usenames,dvipsnames,svgnames,table]{xcolor}

\newcommand{\lfm}{last.fm}

\newcommand{\specialcell}[2][c]{%
  \begin{tabular}[#1]{@{}c@{}}#2\end{tabular}}

\begin{document}

\title{Measuring the Significance of the Geographic Flow of Music}
\author{\IEEEauthorblockN{Conrad Lee,
        Aaron McDaid, and
        P\'{a}draig Cunningham}
        \IEEEauthorblockA{Clique Research Cluster\\
        University College Dublin\\
        8 Belfield Office Park, Clonskeagh\\
        Dublin 4, Ireland\\
        Tel: +353 1 716 5346 \\
        Email: \{conradlee, aaronmcdaid\}@gmail.com, padraig.cunningham@ucd.ie}}

\maketitle

\begin{abstract}
  In previous work, our results suggested that some cities tend to be ahead of
  others in their musical preferences. We concluded that work by noting that to
  properly test this claim, we would try to exploit the leader-follower
  relationships that we identified to make predictions. Here we present the
  results of our predictive evaluation. We find that information on the past
  musical preferences in other cities allows a linear model to improve its
  predictions by approx. 5\% over a simple baseline. This suggests that at best,
  the previously found leader-follower relationships are rather weak.
\end{abstract}

\IEEEpeerreviewmaketitle


\section{Reinterpreting the problem of finding leader-follower relationships as a prediction task}
\begin{figure}[ht!]
  \centering
  \includegraphics[width=0.80\columnwidth]{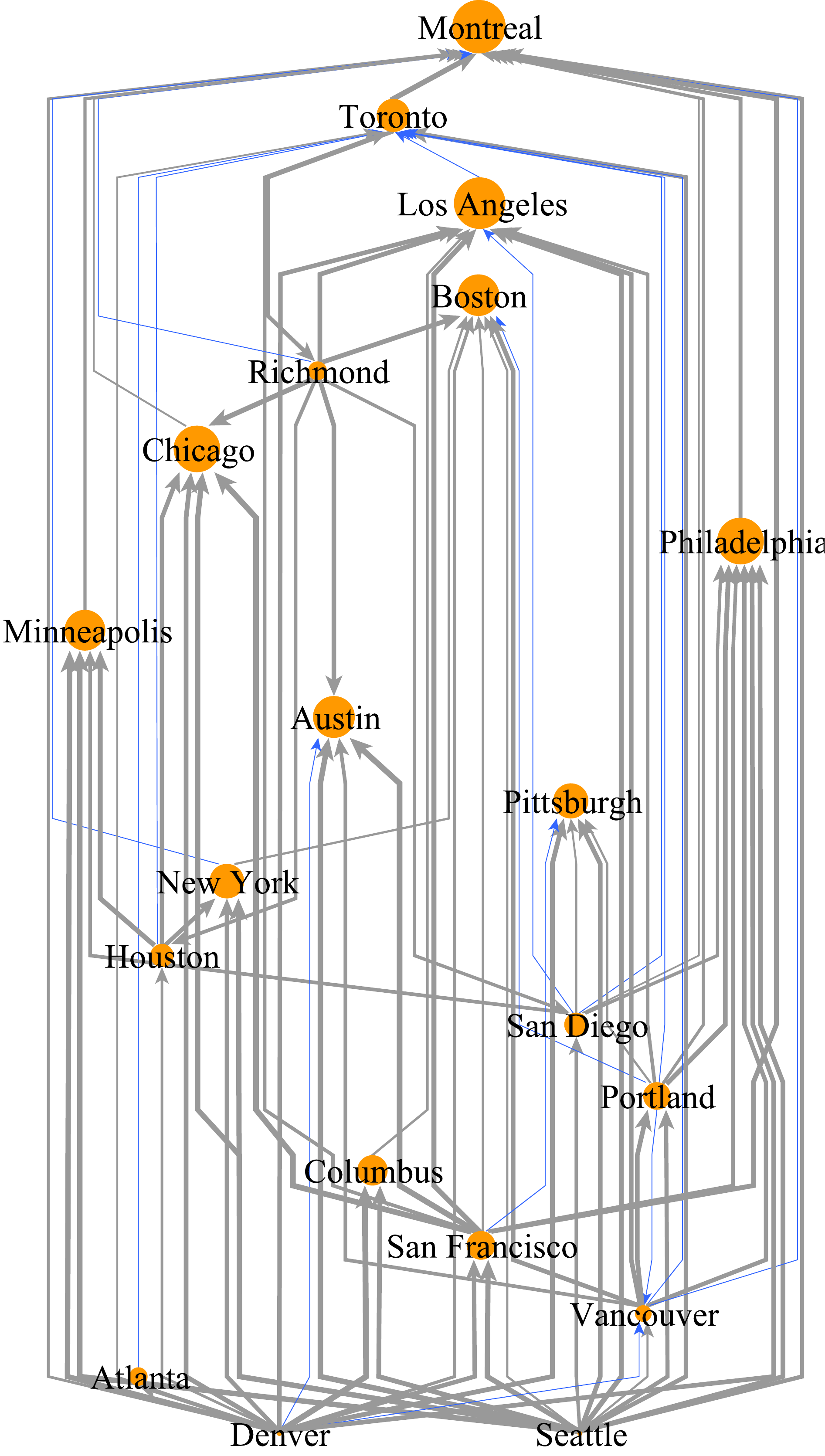}
  \caption{An example of the relationships found in \cite{Lee2012}. This network
    depicts the flow of musical preferences between cities for the genre Indie
  music. See \cite{Lee2012} for details.}
  \label{fig:indie-flow}
\end{figure}

In \cite{Lee2012}, we found that some cities consistently lag others in their
musical preferences. The results (such as those depicted in
\cref{fig:indie-flow}) were surprising, indicating for example that Atlanta,
Montreal, and Oslo are ahead of other cities in their musical preferences. We
were left wondering how \textit{meaningful} the leader-laggard relationships that
we discovered were.

Here we formalize this question in terms of a prediction task. Let us motivate
this prediction task with a simple example: suppose we have observed that Toronto lags
Montreal in indie music by one week, as in \cref{fig:indie-flow}. The relationship
between Montreal and Toronto suggests that,
whenever an indie artist becomes more popular in Montreal, then there should be a
substantially increased probability that the same artist will become more popular
in Toronto one week later.

Given this notion, a straightforward evaluation procedure suggests
itself, based on the following observation: if Montreal really does lead
Toronto, then the present and past information we have about Montreal
should help us to \textit{predict} future music trends in Toronto. The
stronger this relationship is, the better our predictions will be. If,
on the other hand, current information about Montreal only marginally
improves our predictions about Toronto's future, then we can reject our claim
that the relationships between Montreal and Toronto is meaningful.

In the next section, we briefly describe the source of our data and formally
define the change in a city's musical preferences as the velocity of a
city. Next, we specify the prediction task as well as the linear models that we
use for making predictions. Following on, we present the results, which indicate
that the past changes in other cities, when utilized by a linear model, predict
with an accuracy that is approximately 10\% better than predicting that no change
will occur in a city's musical preferences. This finding indicates there is only
a low extent to which music flows from one city to another.

\begin{figure*}[t]
  \centering
  \includegraphics[width=7in]{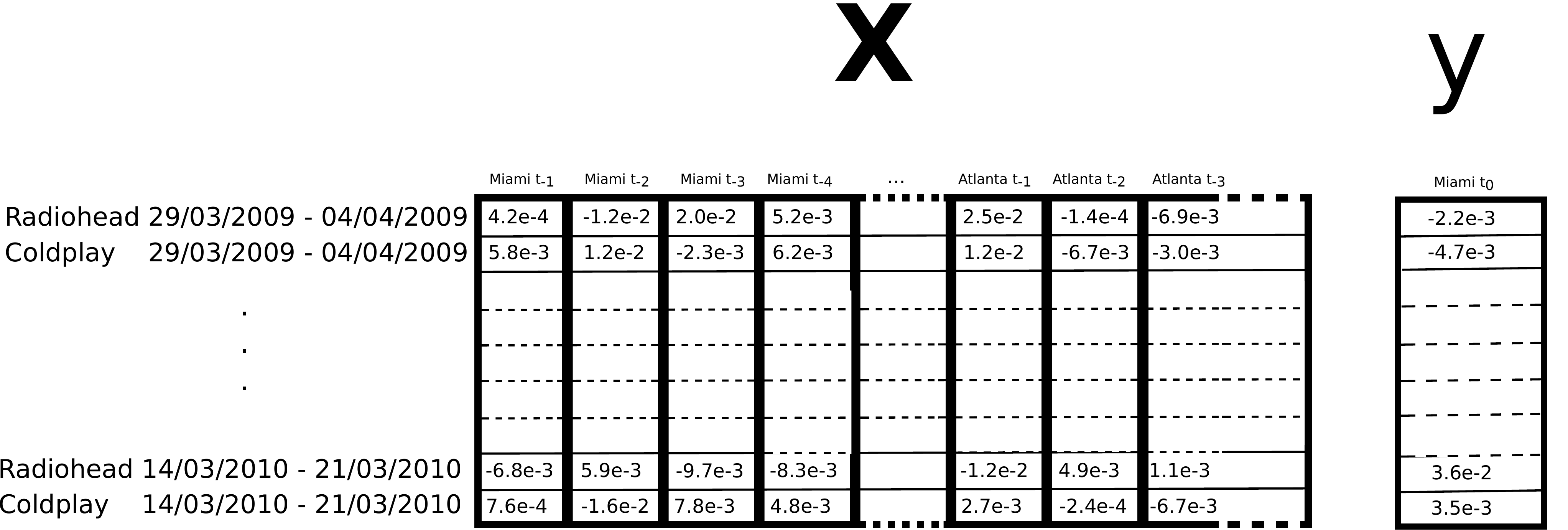}
  \caption{Part of the matrix $\textbf{X}$ used for fitting Miami's linear
    model. In each row, the values of $y$ is the change in that artists'
    popularity in Miami for a given week. The same row in $\textbf{X}$ indicates how that
    artists popularity changed in other cities in the past.}
  \label{music-matrix}
\end{figure*}


\section{Defining a city's veloctiy in artist space}

In this section we provide a brief description of the \lfm{} data, the
preprocessing steps applied to it, and finally motivate and specify the
prediction task mentioned in the previous section.

For each of around 200 cities around the world, \lfm{} publishes a weekly chart
which indicates how many unique listerns each of the five hundred most popular
artists had each week. For a fuller description of \lfm{} and the context of the
data, see Section II of \cite{Lee2012}.

The charts data provided by \lfm{} span every week for more than three years. Let
us label the weeks in order as $t_1, t_2 \ldots t_n$. For each week $t$, we can
create a ``listeners matrix'' such that the rows represent cities, the columns
represent artists, and the values indicate the number of unique listers a given
artist had in a given city. Let us denote the sequence of these listen matrices
as $\mathbf{L}$, and the listeners matrix associated with the $t^{\text{th}}$ week
as $\mathbf{L}_t$. Thus, $\mathbf{L}_{t,c,a}$ is the number of listeners that artist $a$
had in city $c$ in week number $t$. Note that although any single city can have
only 500 non-zero entries in any given week, the matrix as a whole has many
thousands of columns, because accross cities and time, the set of 500 most 
popular artists varies.

A listeners matrix $\mathbf{L}_t$ represents cities as points in a high-dimensional
``artist space.'' If this space is not normalized, then the points which
represent large cities with many users (such as London or New York City) will be
much further away from the origin than cities with fewer listeners. For our
purposes, this is undesirable; instead, we would prefer if two cities that listen
to all artists with the same \textit{proportion} were in the same position in the
artist space. For that reason, in each $\mathbf{L}_t$, we take the Euclidean norm
of each row vector---this ensures that each point's distance from the origin is
one. Let us refer to this normalized version of artist space as
$\mathbf{L}'$--from here on we will only deal with this normalized version of
artist space.

Let us denote the \textit{velocity} $\mathbf{V}_t$ associated with week $t$ to be
$\mathbf{L}'_t$ - $\mathbf{L}'_{t-1}$. Let $\mathbf{V}$ denote the sequence of
$n-1$ consecutive velocity matrices, $\mathbf{V}_2 \ldots \mathbf{V}_{n}$.


\section{Specification of prediction task}

Our prediction task is to estimate the future velocity of some target city. The
results of \cite{Lee2012} indicate that the velocities of some cities
consistently lag those of other cities by between 1 and 8 weeks. This implies
that when trying to predict the future velocity of some city, the past velocities
of \textit{other cities} should help improve our predictions.  If, on the other
hand, our predictive model cannot utilize the past velocities of other cities to
substantially improve these predictions, then we will have evidence that the the
strength of the leader-follower relationship is not sufficient to be useful.

For each city, we will fit two linear models for making predictions, the
\textit{all history model} and the \textit{own history model}. The all history
model will include one coefficient for each (lag size, city) pair and will be fit
on a matrix with the form of $\mathbf{X}$ in \cref{music-matrix}. The own history
model will be the same except it will include only information from the city's
own history; it excludes information on the past velocities of other
cities. Thus, if the all history model fails to outperform the own history
model, then, as stated above, we will conclude that leader-follower relationships
are not very meaningful.

As in \cite{Lee2012}, we will include only the most active US and Canadian cities in
the \lfm{} dataset, and when making a prediction, the model will utilize the
previous eight velocities. Each ``all history model'' model will therefore include 160
coefficients, whereas each ``own history model'' will have eight. Separately, we
perform the same experiment for European cities.

A linear model multplies a matrix $\textbf{X}$ by a vector of coefficients
$\beta$ to produce a vector of predictions, $y$, as in \cref{music-matrix}. We
create two $(\textbf{X}, y)$ pairs: $(\textbf{X}_{\text{train}},
y_{\text{train}})$ and $(\textbf{X}_{\text{test}}, y_{\text{test}})$, where the
former contains the first two years of data, and the latter contains the final
year of data. We use the former to fit the coefficients $\beta$ and the latter to
evaluate the quality of the model.

To measure the error of model, we multiply $\textbf{X}_{\text{test}}$ by $\beta$;
let us call this product $y_{\text{predict}}$. We then take the root-mean squared
error (RMSE) of $y_{\text{test}}$ and $y_{\text{predict}}$. In order to interpret
the size of the RMSE, we compare it to the RMSE of a trivial baseline model,
which predicts that each city's velocity will be zero every week (this would be
the case if musical preferences did not change). The "linear model error" presented in the tables
is in terms of the RMSE of the baseline predictor: a model with 100\% indicates has an error size just as large as
as the baseline's error, 50\% indicates that the model's RMSE is half of the baseline's RMSE.

\begin{table}[t]
\centering
\caption{Prediction accuracy for selected cities in N. America (all)}
\begin{tabular*}{0.90\columnwidth}{lccc}
\toprule 
& \multicolumn{2}{c}{\specialcell{Linear model error \\ (pct baseline)}} &\\
\cmidrule(lr){2-3}
City & Self history & All history & Difference\\
\midrule 
New York & 71.5 & 68.6 & 2.9\\
Phoenix & 78.1 & 74.6 & 3.5\\
\textcolor{Red}{Vancouver} & 78.1 & 74.7 & 3.4\\
\textcolor{OliveGreen}{Pittsburgh} & 78.0 & 74.9 & 3.0\\
Philadelphia & 78.9 & 75.1 & 3.9\\
Minneapolis & 79.3 & 75.1 & 4.2\\
Las Vegas & 77.2 & 75.2 & 2.1\\
\textcolor{OliveGreen}{Atlanta} & 79.8 & 75.4 & 4.5\\
\textcolor{OliveGreen}{Montreal} & 78.3 & 75.6 & 2.7\\
\textcolor{Red}{Denver} & 80.5 & 76.1 & 4.4\\
San Diego & 80.3 & 76.1 & 4.2\\
\textcolor{Red}{Portland} & 80.4 & 76.1 & 4.3\\
\textcolor{OliveGreen}{Houston} & 80.3 & 76.3 & 4.0\\
Columbus & 80.0 & 76.4 & 3.6\\
\textcolor{Red}{Boston} & 80.5 & 76.7 & 3.9\\
Austin & 81.9 & 77.0 & 4.8\\
\textcolor{Red}{San Francisco} & 82.3 & 77.3 & 5.1\\
\textcolor{OliveGreen}{Toronto} & 81.6 & 78.2 & 3.5\\
\textcolor{Red}{Seattle} & 83.5 & 78.5 & 5.0\\
Los Angeles & 83.9 & 78.9 & 5.0\\
\textcolor{OliveGreen}{Chicago} & 84.1 & 79.4 & 4.7\\

\midrule 
\textbf{Avg. all} & 79.9 & 76.0 & 3.8\\
\textcolor{OliveGreen}{\textbf{Avg. leaders}} & & & \textcolor{OliveGreen}{3.7}\\
\textcolor{Red}{\textbf{Avg. followers}} & & & \textcolor{Red}{4.4}\\
\bottomrule
\end{tabular*}
\label{North-America-table-all}
\end{table}


\section{Results}
\label{sec:results}
We run our experiments for all music, and also for a subset of artists who are
classified as ``indie,'' which is a popular genre on \lfm. The results
indicate two points. First, none of the results show large improvement
over the simple baseline predictor that simply predicts a velocity of zero. For
some cities, the RMSE of the ``all history'' is 12\% lower, but that's the maximum
amount of improvement over the baseline, which is not dramatic given that the
baseline is quite trivial. Secondly, we see that the ``all history'' model
outperforms the ``own history'' model, typically achieving twice the improvement
over the baseline.

So we are left to conclude that, in the context of a linear model, information
about past velocities does not allow one to substantially improve predictions. It is true
that a model which includes information about what happens in other cities
performs better than a model which has information about only its own history, but
even with this improvement, our model outperforms the trivial baseline predictor
by only about 10\%.

\begin{table}[t]
\centering
\caption{Prediction accuracy for selected cities in N. America (indie)}
\begin{tabular*}{0.90\columnwidth}{lccc}
\toprule 
& \multicolumn{2}{c}{\specialcell{Linear model error \\ (pct baseline)}} & \\
\cmidrule(lr){2-3}
City & Self history & All history & Difference\\
\midrule 
New York & 72.9 & 70.0 & 2.9\\
Phoenix & 77.8 & 74.1 & 3.7\\
Las Vegas & 76.9 & 74.7 & 2.2\\
Pittsburgh & 79.1 & 75.7 & 3.5\\
Portland & 80.4 & 75.7 & 4.7\\
\textcolor{Red}{Vancouver} & 79.8 & 75.8 & 4.0\\
\textcolor{Red}{Columbus} & 80.1 & 75.8 & 4.3\\
\textcolor{Red}{Denver} & 81.2 & 75.9 & 5.3\\
San Diego & 81.2 & 76.1 & 5.1\\
\textcolor{OliveGreen}{Philadelphia} & 80.3 & 76.3 & 4.0\\
Houston & 81.1 & 76.4 & 4.7\\
\textcolor{Red}{Atlanta} & 81.7 & 76.5 & 5.2\\
Minneapolis & 81.5 & 77.1 & 4.5\\
\textcolor{Red}{Seattle} & 84.1 & 78.0 & 6.1\\
\textcolor{OliveGreen}{Montreal} & 80.7 & 78.4 & 2.3\\
\textcolor{OliveGreen}{Toronto} & 82.8 & 78.6 & 4.1\\
\textcolor{OliveGreen}{Boston} & 82.8 & 78.6 & 4.1\\
Austin & 84.1 & 78.7 & 5.4\\
\textcolor{Red}{San Francisco} & 85.2 & 79.3 & 5.9\\
\textcolor{OliveGreen}{Los Angeles} & 85.8 & 80.5 & 5.3\\
\textcolor{OliveGreen}{Chicago} & 87.2 & 81.0 & 6.2\\

\midrule 
\textbf{Avg. all}  & 81.3 & 76.8 & 4.5\\
\textcolor{OliveGreen}{\textbf{Avg. leaders}} & & & \textcolor{OliveGreen}{4.3}\\
\textcolor{Red}{\textbf{Avg. followers}} & & & \textcolor{Red}{5.1}\\
\bottomrule
\end{tabular*}
\label{North-America-table-indie}
\end{table}

\begin{table}[t]
\centering
\caption{Prediction accuracy for selected cities in Europe (All)}
\begin{tabular*}{0.90\columnwidth}{lccc}
\toprule 
& \multicolumn{2}{c}{\specialcell{Linear model error \\ (pct baseline)}} &\\
\cmidrule(lr){2-3}
City & Self history & All history & Difference\\
\midrule 
\textcolor{OliveGreen}{Dublin} & 70.2 & 67.4 & 2.8\\
\textcolor{Red}{Munich} & 76.9 & 74.1 & 2.8\\
Vienna & 78.3 & 75.5 & 2.8\\
Bristol & 79.9 & 75.6 & 4.3\\
\textcolor{OliveGreen}{Hamburg} & 78.9 & 76.6 & 2.3\\
\textcolor{OliveGreen}{Birmingham} & 80.9 & 77.2 & 3.7\\
\textcolor{OliveGreen}{Leeds} & 81.5 & 77.6 & 3.8\\
Berlin & 80.3 & 77.8 & 2.5\\
\textcolor{Red}{Barcelona} & 81.4 & 77.8 & 3.6\\
\textcolor{Red}{Cracow} & 80.5 & 77.8 & 2.7\\
\textcolor{Red}{Milan} & 80.9 & 78.6 & 2.3\\
\textcolor{Red}{Manchester} & 83.9 & 79.0 & 4.9\\
Madrid & 84.8 & 81.4 & 3.5\\
Paris & 83.8 & 81.5 & 2.3\\
Brighton & 87.7 & 82.6 & 5.1\\
\textcolor{Red}{Warsaw} & 86.2 & 84.1 & 2.1\\
London & 92.5 & 87.8 & 4.7\\
\textcolor{OliveGreen}{Stockholm} & 90.9 & 88.7 & 2.2\\
\textcolor{OliveGreen}{Oslo} & 92.9 & 91.3 & 1.6\\
\midrule 
\textbf{Avg. all}  & 82.8 & 79.6 & 3.2\\
\textcolor{OliveGreen}{\textbf{Avg. leaders}} & & & \textcolor{OliveGreen}{2.7}\\
\textcolor{Red}{\textbf{Avg. followers}} & & & \textcolor{Red}{3.1}\\
\bottomrule
\end{tabular*}
\label{Europe-table-all}
\end{table}

\begin{table}[t]
\centering
\caption{Prediction accuracy for selected cities in Europe (indie)}
\begin{tabular*}{0.90\columnwidth}{lccc}
\toprule 
& \multicolumn{2}{c}{\specialcell{Linear model error \\ (pct baseline)}} &\\
\cmidrule(lr){2-3}
City & Self history & All history & Difference\\
\midrule 
\textcolor{OliveGreen}{Dublin} & 69.9 & 66.7 & 3.2\\
\textcolor{Red}{Bristol} & 78.8 & 74.7 & 4.1\\
\textcolor{Red}{Munich} & 78.6 & 75.2 & 3.5\\
Birmingham & 79.5 & 75.8 & 3.7\\
Vienna & 81.0 & 77.0 & 4.0\\
Manchester & 82.1 & 77.4 & 4.7\\
\textcolor{OliveGreen}{Hamburg} & 80.6 & 77.8 & 2.8\\
Berlin & 81.4 & 78.6 & 2.8\\
\textcolor{Red}{Leeds} & 81.8 & 78.8 & 2.9\\
\textcolor{Red}{Brighton} & 83.7 & 79.8 & 3.9\\
Barcelona & 83.7 & 79.8 & 3.9\\
\textcolor{OliveGreen}{Milan} & 84.7 & 81.6 & 3.1\\
\textcolor{Red}{Cracow} & 86.7 & 82.6 & 4.1\\
\textcolor{OliveGreen}{Paris} & 86.3 & 82.6 & 3.7\\
\textcolor{OliveGreen}{Madrid} & 87.3 & 82.9 & 4.5\\
Stockholm & 90.5 & 87.0 &  3.5\\
London & 91.7 & 87.2 & 4.5\\
\textcolor{Red}{Warsaw} & 93.1 & 89.0 &  4.0\\
\textcolor{OliveGreen}{Oslo} & 93.7 & 91.0 &  2.8\\

\midrule 
\textbf{Avg. all}  & 83.9 & 80.3 & 3.7\\
\textcolor{OliveGreen}{\textbf{Avg. leaders}} & & & \textcolor{OliveGreen}{3.4}\\
\textcolor{Red}{\textbf{Avg. followers}} & & & \textcolor{Red}{3.8}\\
\bottomrule
\end{tabular*}
\label{Europe-table-indie}
\end{table}


\section{Discussion: relating our current findings to those of \cite{Lee2012}}
In \cite{Lee2012}, we found that some cities consistently lag others in their
musical preferences. We were left wondering how \textit{meaningful} the
leader-laggard relationships that we discovered were. Here we have formalized
this question by re-casting our data analysis as a prediction problem. We model a
city's future change in its musical preferences as a linear combination of
previous changes. If the model has access only to its own past velocities, the
model's error is about 80\% of the error associated with a trivial model (predict
a change of zero). When we allow the model to also include the past changes in
other cities, the model's error drops by an additional 3-4\%. These findings
indicate that the previous changes of other cities are only weakly related to any
city's future change in musical preferences.

Let us now consider how our findings here relate to \cite{Lee2012}.  One could
argue that the linear models we use in this work are not comparable to the
methods used in \cite{Lee2012}. Indeed, in that work, there was no explicit
model---we simply adapted a method proposed in \cite{Nagy2010}. However, one
could argue that the network diagrams in \cite{Lee2012} suggest a sort of
``implicit model.'' While it's reasonable to assume that the implicit model is a
linear model, it probably does not allow for negative coefficients. Thus the
linear model we use here (which does allow negative coefficients) is more
expressive than the concepts of flow that we considered in
\cite{Lee2012}. However, even with the more flexible model considered here, we
were not able to make great predictions. This suggests that the model implicit in
\cite{Lee2012} would have been at least as bad, and likely worse, at predicting
changes in musical preferences. In fact, we also tried using linear models whose
coefficients were constrained to be positive, but these models always performed
substantially worse.  Thus, our evaluation here can only put an upper bound on
the quality of the predictions possible with the relationships we proposed in
\cite{Lee2012}.

Can we find any traces of a connection behind our specific findings in
\cite{Lee2012} and our findings here? Are the leader-follower relationships we
identified in that paper not only weak, but also spurious? Ideally we would be
able to compare the leader follower relationships identified in that previous
work with the ones identified here. However, because interpreting
the coefficients of linear models is not straigtforward, there is no clear way
of identifying the leader-follower relationships present in the linear models
developed here.

We can still try to find some indirect connections
between the two sets of results. It is reasonable to assume that for cities which are
laggards, i.e., low down in diagrams like \cref{fig:indie-flow}, information on what
happens in other cities should especially helpful. In other words, if a particular
city is a laggard, then its future changes will tend to be more
determined by the past changes in other cities, whereas if a city is a leader,
then such information will be less useful. Thus laggard cities (highligted in red
in the tables) should benefit more tha leading cities (highlighted in green). In
all four cases (All music and indie music in both N. America and Europe) the
average benefit of the 6 laggard cities was greater than the corresponding
benefit for the leading cities. Thus, it appears that we can make some connection
between our findings here and our findings in \cite{Lee2012}.


\section*{Acknowledgment}
This material is based upon works supported by the Science Foundation Ireland
under Grant No. 08/SRC/I1407: Clique: Graph \& Network Analysis Cluster.


\bibliographystyle{IEEEtran} \bibliography{refs}

\end{document}